\documentstyle[prd, preprint,eqsecnum,aps]{revtex} 
\begin{document}
\draft
 \preprint{\vbox{
\hbox{CTP-TAMU-16/99, SINP-TNP/99-14}
\hbox{hep-th/9905014}
}}

\title{((F, D1), D3) Bound State and
 Its T-dual Daughters\\}

\author {J. X. Lu$^1$ and Shibaji Roy$^2$}
\address{$^1$Center for Theoretical Physics, 
Texas A\&M University, College Station, TX 77843\\
E-mail: jxlu@rainbow.physics.tamu.edu\\
 $^2$Saha Institute of Nuclear Physics,
1/AF Bidhannagar, Calcutta 700 064, India\\E-mail: roy@tnp.saha.ernet.in}

\maketitle

\begin{abstract}
In the previous paper [hep-th/9904129], we constructed a general 
explicit BPS solution for 
(F, D3) non-threshold bound state. By using the SL(2,Z) symmetry of type IIB
string theory, we here construct from (F, D3) a more 
general BPS configuration for a D3 brane with certain units of quantized 5-form
flux and an infinite number of parallel (F, D1)-strings.
  We study its decoupling limit 
and find that given Maldacena's $AdS_5/CFT_4$ correspondence with 
respect to simple D3
branes and with the usual string coupling, we should have a 
similar correspondence
with respect to this 
bound state but now with an effective string coupling. 
We discuss possible descendants of this bound state by T-dualities along 
its longitudinal or transverse directions. In particular, we present 
explicit configurations for ((F, Dp), D(p + 2)) bound states
for $ 2 \le p \le 5$. All these configurations preserve one 
half of the spacetime supersymmetries.
\end{abstract}

\newpage

\section{Introduction\protect\\}
\label{sec:intro}
In the previous paper\cite{lurtwo}, we showed that for the non-threshold 
bound state (F, D3) the dilaton in general does not remain constant as
opposed to 
a simple D3 brane configuration, even though in both cases one half of 
the spacetime supersymmetries are preserved. However, the dilaton was still
 found to be bounded from below by its asymptotic value and from above by 
its finite horizon value. The 
 effective string coupling, defined as $g_{\rm eff} = e^\phi$, 
decreases its value 
 when we move from asymptotic region to the horizon region. This is 
caused by the 
 infinite number of F-strings in the bound state.
In the so-called decoupling limit, this bound state is expected 
to be described either by fields
in the bulk spacetime or by the ${\cal N} = 4$ $U(n)$ SYM theory in 
1 + 3 dimensions (with $n$ the 5-form
charge associated with the 3-brane in (F, D3)). 
In the bulk, we need to study only the near-horizon region under this limit. 
We found that 
the near-horizon string metric of (F, D3) is not automatically 
$AdS_5 \times S^5$ but under
some constant rescalings of the coordinates it becomes $AdS_5 \times S^5$. 
The resulting metric then becomes identical
with the near-horizon geometry of an equivalent simple D3 brane with 
5-form charge $n$ with the 
effective string coupling evaluated at the horizon rather than 
the usual string coupling. 
Following the symmetry argument of
Maldacena \cite{mal}, we mentioned that we must have an $AdS_5/CFT_4$ 
correspondence 
with respect to this `effective' vacuum configuration of the equivalent 
simple D3 brane. The
gauge coupling for the SYM theory becomes related to the effective string 
coupling rather than the
usual string coupling as $g_{\rm YM}^2 = 2 \pi g_{\rm eff}$. 

	After the rescalings of the coordinates, the information 
about the F-strings in the 
(F, D3) in the metric disappeared completely. The only difference 
between
 the equivalent simple D3 brane with the usual 
simple D3 brane just mentioned is the string coupling
 constant. So the effect of the F-strings must be 
encoded into this
 effective string coupling. Indeed, we found that the tension for 
this equivalent simple
 D3 brane is nothing but the tension for the (F, D3) bound state. 
This indicates that the 
`effective' vacuum mentioned above is just the (F, D3) bound state. 
The vacuum configuration
 for the SYM theory in that case does not contain the electric flux lines. 
   We also found  that
 the effect of the electric flux lines in the worldvolume picture of the 
(F, D3) bound state gets 
 absorbed into the gauge coupling constant $g_{\rm YM}^2 = 
2 \pi g_{\rm eff}$. Therefore, we 
 had a consistent picture both in the bulk and on the brane. 
 
Thus the $AdS_5/CFT_4$ correspondence of Maldacena was found to hold true 
even for this non-trivial D3 brane configuration, namely, (F, D3) 
but with
the corresponding effective string coupling rather than the usual 
string coupling. We also observed that
the effective string coupling is quantized in terms of the integral 
charges associated with the F-strings
and the D3 brane in the (F, D3) bound state.  In particular, 
it can be independent of the usual string
coupling in certain limit.

In testing Maldacena's $AdS_5/CFT_4$ conjecture, most of the current activities 
are focused on relating the perturbative particle-like modes of 
type IIB supergravity on
$AdS_5 \times S^5$ with the operators of the boundary ${\cal N} = 4$ SYM
theory. In \cite{lurtwo}, we took a different step in relating 
non-trivial (non-perturbative)
(F, D3) configuration in the bulk to the corresponding one 
in SYM on the boundary.  Actually,
there already exist some efforts \cite{bang,witone,chuhw,liut} 
in this direction,
for example, in \cite{chuhw,liut} certain effects of D-instantons of type
IIB theory in the decoupling limit have been checked to match with those
of the instantons in YM theory on the boundary.

	In this paper, we continue our study for a more general D3 brane 
configuration, 
namely, the
non-threshold ((F, D1), D3) bound state. We  construct this configuration 
explicitly in the following
section by using the type IIB SL(2,Z) symmetry and the (F, D3) configuration 
given in \cite{lurtwo}.
In section 3, we study the properties of this ((F, D1), D3) bound state. In
section 4,  we study  the decoupling limit of this bound state.
We find  that the $AdS_5/CFT_4$ correspondence holds true in the same 
spirit as in \cite{mal} for this
((F, D1), D3) bound states but once again with the corresponding 
effective string coupling rather than
the usual string coupling. In particular, we perform a 
non-trivial check of this
correspondence by showing that the gauge coupling and theta angle obtained 
from the corresponding ${\cal N} = 4$ SYM are indeed related to the effective
string coupling and value of the RR scalar evaluated at the horizon, 
as required 
by the correspondence. The effective string coupling is also quantized 
as in \cite{lurtwo} in terms of integral charges
associated with the F-strings, D-strings and D3 brane in the 
((F, D1), D3) bound state. Interestingly, we
find that D-strings in the ((F, D1), D3) tend to increase the effective 
string coupling while the 
F-strings tend to decrease it. This is also true for the value of the
RR scalar evaluated at the horizon. In section 5, we present all 
possible descendants 
of this ((F, D1), D3) bound state by T-dualities. We also discuss some BPS
states which can be obtained by the type IIB SL(2,Z) acting on the descendants
of the ((F, D1), D3) in the non-perturbative type IIB string theory.
In particular, we 
find non-threshold (Dp, Dp) bound states
 for $1 \le p \le 8$, where the two Dp's share $(p - 1)$ common directions. 
We also find non-threshold (Dp, Dp, Dp) bound states for $1 \le p \le 7$, where
the three Dp's 
share  $(p - 1)$ directions; non-threshold (Dp, Dp, Dp, Dp) bound states for 
$1 \le p \le 6$, where the four Dp's share also $(p - 1)$ 
common directions and
so on.
This can be continued until we reach a seemingly unique non-threshold
(D1, D1, D1, D1, D1, D1, D1, D1, D1) bound state with the nine D1's sharing no 
common direction. All these bound states preserve one half of the spacetime 
supersymmetries.
 We discuss the possible role of these
bound states in the M- or U-theory. We also present the explicit 
configurations for the
((F, Dp), D(p + 2)) bound states for $2 \le p \le 5$. Finally, we conclude 
this paper in section 6.

\section{Non-Threshold ((F, D1), D3) Bound State\protect\\}

In this section, we will construct the 
non-threshold ((F, D1), D3) bound state using the type IIB SL(2,Z) symmetry 
and the (F, D3) 
solution given in \cite{lurtwo}. We will follow the procedure outlined 
in \cite{schone,lurptwo}.  We take the initial configuration as (F, D3) 
solution, 
 with vanishing 
 asymptotic values for all non-vanishing fields.  
 This is given by the following 
 Einstein metric,
 \begin{equation}
 d s^2 = (H' H)^{1/4} \left[ H^{ - 1} \left( - (d x^0)^2 + 
(d x^1)^2\right) + 
 H'^{- 1}\left( (d x^2)^2 + (d x^3)^2 \right) + d y^i d y^i \right],
 \label{eq:d3emz}
 \end{equation}
 with $i = 1, \cdots, 6$;  the dilaton,
 \begin{equation}
 e^\phi = \left(\frac{H'}{H}\right)^{1/2},
 \label{eq:d3dz}
 \end{equation} 
 and the remaining non-vanishing fields,
 \begin{eqnarray}
 H_3^{(1)} &=& - m\,\Delta_{(m, n)}^{- 1/2} \,d H^{- 1}\wedge 
d x^0 \wedge d x^1,
 \nonumber\\
 H_3^{(2)} &=& m\,n \,\Delta_{(m, n)}^{- 1}\,  H'^{ - 2}\, d H 
\wedge d x^2 \wedge d x^3,
 \nonumber\\
 H_5 &=& n \frac{\sqrt{2} \kappa_0 Q_0^3}{\Omega_5} \left(\ast \epsilon_5 + 
 \epsilon_5 \right).
 \label{eq:d3rfz}
 \end{eqnarray}
 In the above, $H^{(1)}_3$ and $H^{(2)}_3$ are  the NSNS and RR 3-form 
field strengths,
  respectively, and 
 they form a doublet 
 \begin{equation}
 {\cal H}_3 =\left(\begin{array}{c}
                 H_3^{(1)}\\
		 H_3^{(2)} \end{array}\right),
\label{eq:fsd}
\end{equation}
under the classical SL(2,R) symmetry of type IIB string theory. 
$H_5$ is the self-dual 
5-form field strength
which is inert under either SL(2,R) or SL(2,Z).  $H$ is a  Harmonic function
 \begin{equation}
 H = 1 + \frac{Q_3}{r^4},
 \label{eq:fhz}
 \end{equation}
 with $r^2 \equiv y^i y^i$ and $Q_3 = \Delta^{1/2}_{(m,n)} 
\sqrt{2} \kappa_0 Q_0^3/ 
 (4 \Omega_5)$, 
 $H'$ is a second Harmonic function defined as,
 \begin{equation}
 H' = 1 + \frac{n^2 Q_3/\Delta_{(m, n)}}{r^4}.
 \label{eq:shz}
 \end{equation}
 Also in the above, the $\Delta$-factor has the form,
 \begin{equation}
 \Delta_{(m,n)} = m^2 + n^2,
 \label{eq:deltaz}
 \end{equation}
 where the asymptotic values of the scalars have been set to zero. Here $n$ 
is the quantized 5-form flux or 
 3-brane charge and
 $m$ is the quantized NS string charge or the number of F-strings through a 
 $ (2\pi)^2 \alpha'$
  area of the 2-dimensional plane perpendicular to the strings in 
(F, D3) as discussed 
 in \cite{lurtwo}.
 $m$ and $n$ are relatively prime integers.
 Unless stated otherwise, $\ast$ always denotes the Hodge dual. 
$\epsilon_n$ denotes 
 the volume form on
 an $n$-sphere and the volume of a unit $n$-sphere is
 \begin{equation}
 \Omega_n = \frac{2 \pi^{(n + 1)/2}}{\Gamma((n + 1)/2)}.
 \label{eq:vns}
 \end{equation}
 In the above $\sqrt{2} \kappa_0 = (2 \pi)^{7/2} \alpha'^2$ and $Q_0^p$ is 
the unit charge for a Dp-brane defined as,
 \begin{equation}
 Q_0^p \equiv (2\pi)^{(7 - 2 p)/2} \alpha'^{(3 - p)/2}.
 \label{eq:cu}
 \end{equation}
 
 It is well-known that type IIB supergravity possesses a classical SL(2,R) 
symmetry \cite{CJ} and a 
 discrete
 subgroup SL(2,Z) of this group is now believed \cite{HT} to survive as 
a quantum symmetry of the 
 non-perturbative type IIB
 string theory. Under a global SL(2,R) transformation $\Lambda$, we have the 
 following
 transformations for the Einstein metric $g_{\mu\nu}$, the 3-form 
field strength 
 doublet ${\cal H}$, the 5-form field strength $H_5$ and the scalar matrix 
 ${\cal M}$ parameterizing 
 the coset SL(2,R)/SO(2) as
 \begin{equation}
 g_{\mu\nu} \rightarrow g_{\mu\nu},~\quad {\cal M} \rightarrow 
\Lambda {\cal M} 
 \Lambda^T,~\quad {\cal
 H}\rightarrow (\Lambda^{-1})^T {\cal H},~\quad H_5 \rightarrow H_5,
 \label{eq:sl2rt}
 \end{equation}
 where the SL(2,R)/SO(2) coset scalar matrix is defined as,
 \begin{equation}
 {\cal M} = e^\phi \left(\begin{array}{cc}
                          e^{- 2 \phi} + \chi^2\qquad&\chi\\
			   \chi \qquad& 1 \end{array} \right),
\label{eq:sm}
\end{equation}
with $\phi$, the dilaton and $\chi$, the axion (or RR scalar) in  
type IIB supergravity.

Since there are an infinite number of (F, D1)-strings in ((F, D1), D3), as 
discussed in 
\cite{lurtwo}, the corresponding charges for F-strings and D-strings 
should be calculated  
according to
\begin{equation}
e^{(i)} = \frac{1}{\sqrt{2} \kappa_0} \int_{R^2 \times S^5} 
\left({\cal M}^{ij} \ast H^{(j)}_3 
- \epsilon^{ij} B^{(j)}_2 \wedge H_5 \right),
\label{eq:dsc}
\end{equation}
where $i, j = 1, 2$ and $\epsilon^{ij}$ is the totally antisymmetric 
SL(2,R) invariant tensor 
with
$\epsilon^{12} = 1$. In general,  $e^{(i)}$ are infinite because 
of the infinite number
of F-strings and D-strings in ((F, D1), D3). As discussed in 
\cite{lurtwo}, the following 
quantities
$Q^{(i)}$ are nevertheless finite and quantized,
\begin{equation}
Q^{(i)} \equiv (2\pi)^2 \alpha' \frac{e^{(i)}}{\sqrt{2} \kappa_0 A_2},
\label{eq:sc}
\end{equation}
where $A_2 = \int d x^2  d x^3$ is the coordinate area of the $x^2x^3$-plane. 
The quantities $Q^{(1)}$ and $Q^{(2)}$ represent charges
 associated 
with the F-strings and D-strings or the number of F-strings and 
D-strings through a  
$(2\pi)^2 \alpha'$
area over the $x^2x^3$-plane measured in some units. From 
Eqs.\ (\ref{eq:dsc}) and (\ref{eq:sl2rt}), we find that
the charge doublet ${\cal Q}^T = (Q^{(1)}, Q^{(2)})$ will transform under 
an SL(2,R) 
transformation as
\begin{equation}
{\cal Q} \rightarrow \Lambda {\cal Q}.
\label{eq:sct}
\end{equation}

We will use the above transformations Eqs.\ (\ref{eq:sl2rt}) 
and (\ref{eq:sct}) to 
obtain the 
non-threshold ((F, D1), D3) configuration from
the given initial (F, D3) solution Eqs.(2.1)--(2.7). Since the 5-form 
field strength is inert under 
the SL(2,R)
transformation, we expect that the expression for $H_5$ given in 
Eq.\ (\ref{eq:d3rfz}) 
remains unchanged.
In other words, the integer $n$ remains the same. However, the 
integral NS-string charge 
$m$ must be
changed under the SL(2,R) transformation $\Lambda$. In general, it 
cannot remain as an integer, 
i.e, quantized.
For this reason, one either needs to introduce a compensating factor 
 for $m$
by hand \cite{schone} or take the initial configuration with an 
arbitrary  classical charge 
$\tilde{\Delta}_{(p,q)}^{1/2}$ \cite{lurptwo} in place of $m$ such that 
the resulting
 NS-string 
and
D-string charges can be quantized. The process of charge quantization 
also determines this factor in terms of 
the quantized
NS-string and RR-string charges and the asymptotic values of the 
dilaton and the axion. 
We will
choose the latter approach, i.e., we take the integer $m$ as an 
arbitrary classical charge  
$\tilde{\Delta}_{(p,q)}^{1/2}$  in the above (F, D3) solution. 

As usual, we start with the zero 
asymptotic 
values of $\phi$ and $\chi$, i.e., ${\cal M}_0 = I$, with $I$ the unit matrix. 
Here 
${\cal M}_0$ denotes
the scalar matrix ${\cal M}$ when the scalars take their asymptotic values. 
We now seek 
a 
$2\times 2$ SL(2,R) matrix 
$\Lambda_0$ which maps the zero asymptotic values of $\phi$ and 
$\chi$ to $\phi_0$ and 
$\chi_0$ (which
are arbitrary but given) as
\begin{equation}
\Lambda_0 I \Lambda_0^T = \Lambda_0 \Lambda_0^T = {\cal M}_0 = 
e^{\phi_0} \left(\begin{array}{cc}
                 e^{-2 \phi_0} + \chi_0^2\qquad& \chi_0\\
		 \chi_0\qquad& 1\end{array}\right).
\end{equation}
The above equation fixes the SL(2,R) matrix $\Lambda_0$ in terms of 
$\phi_0$, $\chi_0$ 
and an undetermined SO(2) angle
$\alpha$ as
\begin{equation}
\Lambda_0 = e^{\phi_0 /2}\left(\begin{array}{cc}
             e^{-\phi_0}\, \cos\,\alpha + \chi_0\,\sin\,\alpha 
\qquad& - e^{-\phi_0}\,\sin\,\alpha \\
	     \sin\,\alpha \qquad& \cos\,\alpha \end{array}\right).
\label{eq:sl2rm0}
\end{equation}

Imposing charge quantization on $Q^{(i)}$  as $Q^{(1)} = p, Q^{(2)} = q$ 
with $p, q$ relatively prime integers for our general ((F, D1), D3) 
bound state, we  
have from 
Eq.\ (\ref{eq:sct}) 
\begin{equation}
\left(\begin{array}{c}
      p\\
      q \end{array} \right) = \Lambda_0 \left(\begin{array}{c}
     \tilde{\Delta}_{(p,q)}^{1/2} \\
      0 \end{array} \right).
\label{eq:qsct}
\end{equation}
The above equation fixes not only the SL(2,R) matrix $\Lambda_0$ 
completely but also the
$\Delta_{(p,q)}$-factor in terms of $p, q$ and $\phi_0, \chi_0$ as
\begin{equation}
\Lambda_0 = \frac{1}{\tilde{\Delta}_{(p,q)}^{1/2}} \left(\begin{array}{cc}
p \qquad& -  q \,e^{-\phi_0} + \chi_0 \,(p - \chi_0 q) \,e^{\phi_0} \\
q\qquad & (p - \chi_0 q)\, e^{\phi_0} \end{array} \right),
\label{eq:csl2rm0}
\end{equation}
and
\begin{equation}
\tilde{\Delta}_{(p, q)} = e^{\phi_0}\, (p - \chi_0 \,q)^2 + e^{-\phi_0} q^2,
\label{eq:deltaf}
\end{equation}
which is invariant under an SL(2,Z) transformation. Note that in this process
the SO(2) angle $\alpha$ gets fixed as $e^{i\alpha} = \left[(p - \chi_0 q)
e^{\phi_0/2} + i q e^{-\phi_0/2}\right] \tilde{\Delta}_{(p,q)}^{-1/2}$.

Once we have the SL(2,R) matrix $\Lambda_0$ (Eq.\ (\ref{eq:csl2rm0})), 
we can 
obtain a general 
((F, D1), D3) configuration  by performing the SL(2,R) transformation as 
given in
Eq.\ (\ref{eq:sl2rt}) with $\Lambda = \Lambda_0$ on the initial (F, D3) 
configuration given in Eqs.(2.1)--(2.7),
with $m$ replaced by $\tilde{\Delta}_{(p,q)}^{1/2}$. This 
configuration is
characterized in terms of the quantized charges $p$ and $q$ associated 
with the 
F-strings and D-strings, the quantized charge $n$ of the D3 brane, 
and arbitrary but given 
asymptotic values $\phi_0$ and $\chi_0$ of the dilaton 
$\phi$ and the axion $\chi$, respectively. Thus, we have 
the following
explicit configuration of the bound state ((F, D1), D3). The 5-form 
$H_5$ remains unchanged, as given in 
Eq.\ (\ref{eq:d3rfz}).
 The Einstein metric $d s^2$ and the Harmonic functions $H$ and $H'$ all 
remain the same
in forms as before, but the $\Delta_{(m, n)}$-factor given in 
Eq.\ (\ref{eq:deltaz}), 
which appears in the
Harmonic functions, is changed to
\begin{eqnarray}
\Delta_{(p, q, n)} &=& \tilde{\Delta}_{(p,q)} + n^2,\nonumber\\
                   & = &  (p -\chi_0\, q)^2 \, e^{\phi_0} + 
q^2\, e^{-\phi_0} + n^2,
\label{eq:ndelta}
\end{eqnarray}
which is also invariant under SL(2,Z) as expected since $n$ is 
inert under SL(2,Z).
The dilaton is now
\begin{equation}
e^\phi = e^{\phi_0} \frac{H''}{\sqrt {H\,H'}},
\label{eq:ndilaton}
\end{equation}
and the axion is
\begin{equation}
\chi = \frac{\chi_0 H' + (H - H') p q e^{- \phi_0} / 
\tilde{\Delta}_{(p,q)}}{H''}.
\label{eq:naxion}
\end{equation}
In the above, we have introduced a third Harmonic function $H''$ as
\begin{equation}
H'' = 1 + \frac{(n^2 + q^2 \, e^{-\phi_0}) Q_3 /\Delta_{(p, q, n)}}{r^4},
\label{eq:thirdhf}
\end{equation}
where $Q_3 = \Delta^{1/2}_{(p, q ,n)} \sqrt{2} \kappa_0 Q_0^3/ 
(4 \Omega_5)$. As 
expected,
both the dilaton and the axion approach their respective asymptotic values as 
$r \rightarrow \infty$.
Finally, the NSNS and RR 3-form field strengths are
\begin{eqnarray}
H_3^{(1)} &=& - e^{\phi_0}\,(p - \chi_0\, q)\, 
\Delta^{-1/2}_{(p, q, n)} d H^{- 1} 
\wedge d x^0 \wedge d x^1
            - q\, n \, \Delta^{ - 1}_{(p, q, n)} H'^{-2} 
d H\wedge d x^2 \wedge d x^3,
	    \nonumber\\
H_3^{(2)} &=& \left[ \chi_0\, (p - \chi_0\, q) e^{\phi_0} - q \, 
e^{-\phi_0} \right] 
\Delta^{-1/2}_{(p, q, n)}
             d H^{-1} \wedge d x^0 \wedge d x^1 \nonumber\\
	   &\,&\qquad  + \,p\, n \Delta^{ - 1}_{(p, q, n)} 
	     H'^{-2} d H \wedge d x^2 \wedge d x^3.
\label{eq:3ffs}
\end{eqnarray}

In the following section, we will  study 
certain properties of this configuration.

\section{Properties of ((F, D1), D3) Bound State \protect\\}
\label{sec:p}

By construction, we know that the configuration of ((F, D1), D3) given in 
the previous section carries
one $(p,q)$-string along the $x^1$-axis per $(2\pi)^2 \alpha'$ area over the
$x^2x^3$-plane. There are $p$ F-strings and $q$ D-strings in the 
$(p,q)$-string.
 This non-threshold bound state also carries the 5-form charge $n$.

We would like to mention that in constructing the SL(2,Z) invariant
((F, D1), D3) bound state solution of type IIB theory, we have chosen
the Einstein-frame metric to be asymptotically Minskowski. 
However, since we are here studying strings and Dp branes in string theory, 
it 
would be natural to choose the string-frame metric to be asymptotically 
Minkowski instead. Moreover,
we need to calculate the tension for this bound state and compare it with 
the corresponding worldvolume result where the string-frame metric is 
always chosen
to be asymptotically Minkowski in the context of string theory. 
Also, usually in discussing the so-called decoupling limit,  
the string-frame metric is chosen to be asymptotically Minkowski.  
For these reasons, we re-express the explicit solution given in the
previous section in terms of  such a choice for the asymptotic 
metric\footnote{Actually
there are some rationality behind the choice of the string-frame metric 
to be asymptotically
Minkowski and relating such a choice to the SL(2,Z) symmetry of 
type IIB theory.
We will give a detail account of this along with other things 
in a separate paper.}.
For this choice the 5-form field strength continues to be given by 
the expression in 
Eq.\ (\ref{eq:d3rfz}). The dilaton and the axion also remain the same as 
given by Eqs.\ (\ref{eq:ndilaton}) and \ (\ref{eq:naxion}), respectively. 
The same is true for
the Harmonic functions $H$, $H'$ and $H''$ but with
the replacements $Q_3 \rightarrow e^{3 \phi_0 /2} Q_3$, 
$ n^2\, Q_3/\Delta_{(p,q,n)} \rightarrow n^2 e^{\phi_0/2} Q_3/
\Delta_{(p,q,n)}$,
and $ (n^2 + q^2\, e^{- \phi_0}) Q_3/\Delta_{(p,q,n)} 
\rightarrow (n^2 + q^2) e^{\phi_0/2} Q_3/\Delta_{(p,q,n)}$ 
in the respective Harmonic
functions. Here the $\Delta_{(p,q,n)}$ is also changed to
\begin{equation}
\Delta_{(p, q, n)} = (p - \chi_0\, q)^2\, e^{\phi_0} + (q^2 + n^2)\, 
e^{- \phi_0},
\label{eq:nndelta}
\end{equation}
which is not manifestly SL(2,Z) invariant now.
Apart from an overall constant  factor $e^{- \phi_0/2}$ and the changes in the
corresponding Harmonic functions $H$ and $H'$ mentioned above, 
the Einstein-frame
metric Eq.(2.1) remains the same in form. The NSNS and RR 3-form 
field strengths are now changed to
\begin{eqnarray}
H_3^{(1)} &=& - e^{\phi_0/2}\,(p - \chi_0\, q)\, 
\Delta^{-1/2}_{(p, q, n)} d H^{- 1} 
\wedge d x^0 \wedge d x^1
            - e^{- \phi_0} q\, n \, \Delta^{ - 1}_{(p, q, n)} 
H'^{-2} d H\wedge d x^2 \wedge d x^3,
	    \nonumber\\
H_3^{(2)} &=& e^{- \phi_0/2} \left[ \chi_0\, (p - \chi_0\, q) 
e^{\phi_0} - q \, e^{-\phi_0} \right] 
\Delta^{-1/2}_{(p, q, n)}
             d H^{-1} \wedge d x^0 \wedge d x^1 \nonumber\\
	   &\,&\qquad + e^{- \phi_0}p\, n \Delta^{ - 1}_{(p, q, n)} 
	     H'^{-2} d H \wedge d x^2 \wedge d x^3,
\label{eq:n3ffs}
\end{eqnarray}
where $\Delta_{(p, q, n)}$ is given by Eq.\ (\ref{eq:nndelta}).

We can now calculate the mass per unit 3-brane volume for this 
bound state with the new 
Einstein-frame metric using the generalized formula given in 
\cite{lurtwo}. We therefore
find the tension for the ((F, D1), D3) bound state as
\begin{equation}
T_3 (p, q, n) = \frac{T_0^3}{g} \sqrt{n^2 + q^2 + g^2\, (p - \chi_0 q)^2},
\label{eq:d3tension}
\end{equation}
which, for $\chi_0 = 0$, agrees precisely with what we 
obtained from the worldvolume study 
in \cite{lurone}.

The expression for tension in (3.3) clearly shows that the spacetime 
((F, D1), D3) bound state should be 
identified with the bound state consisting of D3 branes carrying  
quantized constant
electric and magnetic fields given in \cite{lurone} based on 
the worldvolume study. In other words, the
$(p,q)$-strings should be identified with the quantized electric and 
magnetic flux lines. The
quantized constant electric and magnetic fields are related to each other 
by the 
SL(2,Z) symmetry in the ${\cal N} = 4$ SYM theory on the worldvolume. 
This is directly 
linked to the type IIB SL(2,Z) relating F-strings and D-strings in the 
$(p,q)$-string provided
we identify the gauge coupling and the $\theta$-angle in the gauge 
theory with the string
coupling and the asymptotic value of the axion as 
\begin{equation}
\frac{2 \pi i}{g_{\rm YM}^2} +  \frac{\theta}{4\pi} = 
\frac{i}{ g} +  \chi_0,
\label{eq:cr}
\end{equation}
where $g = e^{\phi_0}$ is the string coupling\footnote{Our
convention for the complex gauge coupling follows from the Born-Infeld action
which differs from the standard one for $SU(n)$ theories by a factor 2. 
We thank Juan Maldacena for discussion on this issue.}.  

\section{The Decoupling Limit of ((F, D1), D3) Bound State\protect\\}
\label{sec:dl}

Let us now study the decoupling limit for this ((F, D1), D3) bound state. Under
this limit, 
\begin{equation}
g^2_{\rm YM} = 2 \pi g = {\rm fixed},~\quad U = \frac{r}{\alpha'} = {\rm fixed},
~\quad \alpha' \rightarrow
0,
\label{eq:dcl}
\end{equation}
we know that the modes propagating on the 3-brane worldvolume 
decouple from the modes
propagating in the bulk spacetime. Also since $\alpha' \rightarrow 0$ 
is a low-energy limit,
the modes propagating on the brane are just the massless ones of the 
open strings.
Therefore, D3 brane is described by the ${\cal N} = 4$ SYM theory. 
It is also described
by the fields in the bulk. So we have two equivalent descriptions 
for the D3 brane.

{}From the spacetime point of view, the decoupling limit tells us that 
we need to study
only the near-horizon geometry. Under this limit, the dilaton, 
which defines the effective 
string coupling $g_{\rm eff}$, is given as,
\begin{equation}
g_{\rm eff} \equiv e^\phi = e^{\phi_0} \frac{1 + q^2/n^2}{\left(1 + [q^2 + 
e^{2 \phi_0} (p - \chi_0 q)^2]/n^2\right)^{1/2}},
\label{eq:esc}
\end{equation}
and the string-frame metric, in terms of this $g_{\rm eff}$, is
\begin{eqnarray}
d s^2 = &&\alpha' \left[ \frac{U^2}{\sqrt{4 \pi n g_{\rm eff}}}
\frac{n^2 + q^2}{n^2 + q^2 + e^{2\phi_0}
(p - \chi_0 q)^2} \left( - (d x^0)^2 + (d x^1)^2 \right) \right. \nonumber\\
&&\qquad + \frac{U^2}{\sqrt{4 \pi n g_{\rm eff}}} (1 + q^2 /n^2) 
\left( (d x^2)^2 + ( d x^3)^2 \right)
\nonumber\\
&&\qquad \left.  + \sqrt{4 \pi n  g_{\rm eff}} \left( \frac{d U^2}{U^2} 
+ d \Omega_5^2\right)\right].
\label{eq:nhsm}
\end{eqnarray}
We are looking for an equivalent simple D3-brane description 
for this ((F, D1), D3)
in the decoupling limit. We can achieve this minimally by rescaling
$x^0$ and $x^1$ such that we end up with
\begin{eqnarray}
d s^2 =\alpha' &&\left[\frac{U^2}{\sqrt{4 \pi n g_{\rm eff}}} (1 + q^2 /n^2) 
\left( - (d x^0)^2 + (d x^1)^2 + (d x^2)^2 + ( d x^3)^2 \right)\right. 
\nonumber\\
 &&\qquad\left. + \sqrt{4 \pi n  g_{\rm eff}} \left( \frac{d U^2}{U^2} 
+ d \Omega_5^2\right)\right].
\label{eq:nhadsm}
\end{eqnarray}
This metric has the isometries of $AdS_5 \times S^5$ but not in the
standard form because of the factor $(1 + q^2/n^2)$.  
The metric in (4.4) therefore describes
the near-horizon geometry of a simple D3 brane 
 with 5-form flux
$n$ but with the string coupling given by $g_{\rm eff}$ rather 
than the usual string coupling $g =
e^{\phi_0}$. The extra factor $(1 + q^2/n^2)$ in the above metric
indicates that unlike the F-strings, the presence of D-strings in D3 brane
worldvolume changes the entire D3 brane worldvolume metric. This factor 
does not
play an important role in the near-horizon metric 
Eq.\ (\ref{eq:nhadsm}) but it does
have influence on the asymptotic metric of the equivalent 
simple D3 brane as we will
explain below. In the decoupling limit, D-brane picture implies that 
this equivalent simple  D3 brane is also 
described by the 
${\cal N} = 4$  $U(n)$ SYM theory in 1 + 3 dimensions with the gauge coupling 
$g_{\rm YM}^2 = 2 \pi g_{\rm eff}$ and the theta angle  
$\theta = 4 \pi \chi_{\rm eff}$ with $\chi_{\rm eff}$ 
evaluated at the horizon as
\begin{equation}
\chi_{\rm eff} =  (pq + n^2 \chi_0)/(n^2 + q^2).
\label{eq:chi}
\end{equation}
Thus in the 
same spirit of Maldacena,
we should have an $AdS_5/CFT_4$ correspondence here but with the 
string coupling $g_{\rm eff}$ and the
axion $\chi_{\rm eff}$. 

This equivalent simple D3 brane is taken as the ``effective" vacuum 
configuration for this
correspondence. In analogy with the study of the decoupling limit of 
the (F, D3) bound state  in
\cite{lurtwo}, we may anticipate that this effective
vacuum corresponds to ((F, D1), D3) bound state. In order to provide some 
evidence for this we notice that once the above
coordinate rescalings are done, the information about the (F, D1) bound 
states in the ((F, D1), D3) disappears
in the resulting metric. However, unlike the case for (F, D3) discussed in 
\cite{lurtwo}, we have the extra factor $(1 + q^2/n^2)$ appearing 
in the metric and therefore
we should be careful in identifying $g_{\rm eff}$ as the string coupling in the
equivalent D3 brane. We discuss in the following how to make this 
identification in an unambiguous manner.

	Note that in order to find a stable BPS configuration of D3 branes 
(in general $p$-branes)
from the corresponding supergravity, we usually insist either 
the Einstein frame or the
string frame metric to be asymptotically Minkowski. There actually 
exists a general
choice for the asymptotic metric which includes the above two choices. 
Specifically,
 we can re-obtain the $p$-brane solutions in \cite{dufkl} by insisting
 the Einstein metric $g_{MN}$ to go to  $e^{V_0} \eta_{MN}$
asymptotically with $V_0$ an arbitrary but given constant and 
$\eta_{MN}$ the flat
Minkowski metric. For example,  $V_0 = 0$ corresponds to choosing the Einstein
metric to be asymptotically Minkowski while $V_0 = - \phi_0/2$ corresponds 
to choosing 
the string metric to be asymptotically Minkowski with $\phi_0$ 
the asymptotic value of the dilaton.
Let us specialize for the D3 brane in type IIB theory. With this choice 
of asymptotic 
Einstein metric, we can repeat the calculation in \cite{dufl} 
to have the following
Einstein metric 
\begin{equation}
d s_E^2 = e^{V_0} \big[ H^{- 1/2} (- (d x^0)^2 +
\cdots + (d x^3)^2 ) + H^{ 1/2} dy^i
dy^i \big]
\label{eq:sd3m}
\end{equation}
where the Harmonic function $H$ is now
\begin{equation}
H = 1 + \frac{4 \pi \alpha'^2 n }{r^4} e^{ - 2 V_0},
\end{equation}
and the integer $n$ is the 5-form flux.
If we denote the constant dilaton for this solution as $c_0$, 
then the string frame 
metric is
\begin{equation}
d s^2 = e^{c_0 /2 + V_0} \big[ H^{- 1/2} (- (d x^0)^2 +
\cdots + (d x^3)^2 ) + H^{ 1/2} dy^i
dy^i \big],
\label{eq:sfm}
\end{equation}
 As usual, in obtaining this metric we have set the brane 
$\sigma$-model coordinates
 $\sigma^\mu = x^\mu$ for $\mu = 0, 1, 2, 3$. 
 
 Now let us examine the near-horizon behavior of the above metric 
in the decoupling limit. 
 We have ($U = \frac{r}{\alpha'} = {\rm fixed}$ as $\alpha' \rightarrow 0$) 
 \begin{equation}
 d s^2 =  \alpha' \left[ \frac{U^2}{\sqrt{4 \pi n g_s}} 
g _s\,e^{ 2 V_0} \left(- (d x^0)^2
 + \cdots + (d x^3)^2 \right) + \sqrt{4 \pi n g_s} 
\left( \frac{d U^2}{U^2} + d \Omega_5^2\right)
 \right].
 \label{eq:esd3m}
 \end{equation}
 where $g_s = e^{c_0}$ is the present string coupling. We note from (4.9) 
that: 1) The
 coefficient in front of $d U^2 / U^2$ or $d \Omega_5^2$ is always given by 
 $\sqrt{4 \pi n g_s}$, independent of the constant $V_0$. This provides 
us a way
 to read off the string coupling unambiguously from a near-horizon 
geometry of D3 brane.
 2) Also, unless $V_0 = - c_0/2$ (i.e., the string metric is 
asymptotically Minkowski), 
 the near-horizon metric of a D3 brane is in general not 
in the standard form of 
 $AdS_5 \times S^5$. Now by comparing Eq.\ (\ref{eq:nhadsm}) 
with (\ref{eq:esd3m}),
 we can see that the $g_{\rm eff}$ is indeed the string
 coupling for the equivalent simple D3 brane and the 
asymptotic string metric for the 
 equivalent D3 brane is not Minkowski but  
 \begin{equation}
 e^{V_0}\eta_{MN} = g_{\rm eff}^{ -1/2} (1 + q^2/n^2)^{1/2}\eta_{MN}.
 \label{eq:v0}
 \end{equation}

We also note from Eqs.(4.4) and (4.9) that the differences between the
equivalent D3 branes and the usual D3 branes (with string coupling $g =
e^{\phi_0}$ and $V_0 = - \phi_0/2$) are in the string coupling constant and
the asymptotic form of the metric.
 So the effects of these (F, D1) 
bound states must be encoded in
the effective string coupling $g_{\rm eff}$ given in Eq.\ (\ref{eq:esc}) 
and the factor 
$1 + q^2/n^2$. First let us examine the tension for the equivalent D3 brane. 
Since we have already
determined its string coupling as $g_{\rm eff}$, 
so the tension for this equivalent 
simple D3 brane is $n T_0^3 / g_{\rm eff}$. Now if we use the explicit 
expression for the $g_{\rm eff}$, we find
\begin{equation}
\frac{n T_0^3}{g_{\rm eff}} = \frac{T^3_0}{g} \sqrt{n ^2 + q^2 + g^2\, 
(p - \chi_0 q)^2}\, \left( 1 +
q^2/n^2\right)^{-1}.
\label{eq:etension}
\end{equation}

Comparing it with Eq. (3.3), we find that they differ by a factor 
$(1 + q^2/n^2)^{-1}$. 
Recall that in obtaining (3.3), we calculated the tension against 
asymptotic D3
worldvolume whose metric is asymptotically Minkowski. In other words, 
the worldvolme is 
calculated according to $A_4 = \int d^4 \sigma$. However, the tension 
in Eq.\ (\ref{eq:etension}) for the equivalent D3 brane 
is calculated with respect to its own
asymptotic string metric which is in general not the same as the 
string metric for the
original ((F, D1), D3) configuration. In the present case, the 
asymptotic string metric 
for the equivalent D3 brane is $ g^{1/2}_{\rm eff} e^{V_0} \eta_{MN}$ 
with $e^{V_0}$ given
by Eq.\ (\ref{eq:v0}). For $M, N = 0, 1, 2, 3$, it gives the 
worldvolume asymptotic
metric. Therefore, the tension in Eq.\ (\ref{eq:etension}) is 
calculated against the
worldvolume $A'_4 = g_{\rm eff} e^{2 V_0} \int d^4 \sigma$. 
In order to make a proper
comparison with the tension (3.3), the present tension should be 
measured with respect to
the same worldvolume, i.e., $A_4$ rather than $A'_4$. In other words, 
we expect that 
\begin{equation}
\frac{n T_0^3}{g_{\rm eff}} \frac{A'_4}{A_4}
\end{equation}
should be the same as the tension of ((F, D1), D3) given by Eq. (3.3). 
This is indeed true
since $A'_4 / A_4 = (1 + q^2/n^2)$. This indicates 
 that the vacuum configuration for the $AdS_5/CFT_4$ 
correspondence in this case is 
the ((F, D1), D3) bound state.

 On the SYM side, the $AdS_5/CFT_4$ correspondence suggested above 
implies that the electric and magnetic
flux lines should disappear, too. As claimed, the gauge coupling 
and the theta angle
should now be given as 
$g_{\rm YM}^2 = 2 \pi g_{\rm eff}$ and $\theta = 4 \pi \chi_{\rm eff}$, 
respectively.
Showing these two relations from the SYM side independently is 
certainly a non-trivial
check of this correspondence.  We will show below that 
these two relations are
 indeed true from the SYM side.

We know from D-brane picture that the dynamics of a single D3 brane 
(or $n$ coincident D3 branes) is 
described in general, in the low energy limit, by the abelian 
(or non-abelian $U(n)$) Born-Infeld action 
of ${\cal N} = 4$ SYM coupled with background fields. In the 
decoupling limit, these background
fields are decoupled and therefore the D3 brane (or $n$ coincident D3 branes) 
is described purely by the 
gauge fields with flat Minkowski background. Since the non-abelian version 
of the Born-Infeld action
for $n$ coincident D3 branes is not known yet\footnote{A possible form was
suggested by Tseylin \cite{tse}.}, we limit ourselves here 
to the abelian Born-Infeld
action of a single D3 brane.  This turns out to be sufficient for confirming 
the abovementioned relations.

Our worldvolume study in \cite{lurone} and the spacetime study in 
\cite{lurtwo} show that the F-strings and
D-strings in ((F, D1), D3) correspond to the quantized constant electric field 
$ (2 \pi \alpha') E = g (p - q \chi_0)$ and the quantized constant 
magnatic field
$(2 \pi \alpha') B = q$, respectively, in the linear 
approximation\footnote{Even though
these electric and magnetic fields are good only in the linear approximation, 
the tension obtained
in \cite{lurone} is actually exact. This is the property of a BPS state.}. 
 For a single D3 brane
(i.e., $n = 1$), while the linear approximation can be justified 
for the electric field $E$ 
by insisting a small string coupling $g$, this approximation is not valid, 
strictly speaking, for 
the magnetic field $B$ with a non-vanishing integer $q$. Even so, we 
obtain the correct
tension for the bound state ((F, D1), D3). The success in obtaining 
the tension for 
((F, D1), D3) even for $n \neq 1$ seems to indicate that we may take the 
$n$ coincident D3 branes
effectively as a single D3 brane  with a tension $n T_0^3$ and the 
constant electric field as
$(2 \pi \alpha') E = g (p - q \chi_0)/n$ and the constant magnetic field as 
$(2 \pi \alpha') B = q/n$ at least for these two constant background fields. 
If we assume this, 
the linear approximation can always be made by insisting, for 
example, $n >> q, n >> p$, with
small $g$. We will take these effective $E$  and $B$ as our constant 
background fields where
$n = 1$ is just a special case.

Let us recall that in \cite{lurone}, we picked a special direction, 
i.e, the $x^1$-direction, as the direction for
the quantized constant electric and magnetic fields. By doing so, 
we break the Lorentz $SO(1,3)$ to
$SO(1,1) \times SO(2)$. In order to find the corresponding description 
of the bulk equivalent simple 
D3 brane, we now need to make a consistent ansatz for the background 
gauge fields which
respects the $SO(1,3)$ symmetry as well as the translational symmetry.
This can be achieved in the following two ways. We can take a stochastic
averaging of a constant homogeneous background as in \cite{liut}. Or we can
follow \cite{min} to take the constant homogeneous background as the 
ground-state expectation value. We here follow the latter approach and have
\begin{eqnarray}
&&< {\cal F}_{\mu \nu} > = 0,\nonumber\\
&& (2 \pi \alpha')^2 < {\cal F}_{\mu \nu} {\cal F}^{\lambda \sigma} > 
= \frac{q^2 - g^2 (p - q
\chi_0)^2}{6 n^2}\, \delta_{\mu\nu}^{\lambda \sigma} - 
\frac{g q (p - q\chi_0)}{3 n^2} \,
\epsilon_{\mu\nu}\,^{\lambda\sigma},
\label{eq:ansatz}
\end{eqnarray}
where $<\cdots >$ denotes the vacuum expectation value of field operators, 
$\delta_{\mu\nu}^{\lambda\sigma} = \delta_\mu^\lambda 
\delta_\nu^\sigma - \delta_\nu^\lambda
\delta_\mu^\sigma$ and $\epsilon^{\mu\nu\lambda\sigma}$ is the totally 
antisymetric tensor with
$\epsilon^{0123} = 1$ ($\mu, \nu, \lambda, \sigma = 0, 1, 2, 3$). 
In the above, we first have $ < {\cal F}^2 > = 2 a$ and 
$< {\cal F} \tilde{{\cal F}}> = 4 b $ with $a$ and $b$ the corresponding constant 
VEV's\footnote{Non-vanishing 
$< {\cal F} \tilde{{\cal F}}>$ implies the spontaneous breaking of P and CP
invariances.}. Then we determine the right
side of the above second equation by insisting $ a = (B^2 - E^2)$ and  
$b = E B $
with $E$ and $B$ the background fields given before. Also, here 
${\tilde {\cal F}}_{\mu \nu} = (1/2) \epsilon_{\mu\nu}\,^{\lambda\sigma} 
{\cal F}_{\lambda\sigma}$.

Since our purpose is to find the effective gauge coupling and 
the effective theta angle for the
fluctuations of gauge fields with respect to the above background, 
so we need to consider only the gauge
fields in the BI action with the Lagrangian 
\begin{equation}
{\cal L} = - \frac{T_3}{g} \sqrt{- {\rm det}(\eta_{\mu\nu} + 
(2 \pi \alpha') {\cal F}_{\mu\nu})} 
             + \frac{\chi_0 T_3}{8} (2 \pi \alpha')^2 
\epsilon^{\mu\nu\lambda\sigma} {\cal F}_{\mu\nu}
	     {\cal F}_{\lambda\sigma},
\label{eq:bia}
\end{equation}
where $g = e^{\phi_0}$ and $\chi_0$ are the asymptotic string coupling 
and the asymptotic value of the
RR scalar, respectively, and $T_3 = n T_0^3$.

Now we write ${\cal F}_{\mu\nu} = f_{\mu\nu} + F_{\mu\nu}$ with $f_{\mu\nu}$, 
the fluctuating field and
$F_{\mu\nu}$, the background, i.e., $< f_{\mu\nu}> = 0$ and 
$<f_{\mu\nu} f^{\lambda\sigma}> = 0$ and
expand the above Lagrangian, with respect to the background, to quadratic order
in $f_{\mu\nu}$. 
We end up with a quadratic Lagrangian of $f_{\mu\nu}$ as
\begin{equation}
{\cal L}_f = \frac{1}{8} < \frac{\partial^2 {\cal L}}
{\partial {\cal F}_{\mu\nu} 
\partial {\cal F}^{\lambda\sigma}} > f_{\mu \nu} f^{\lambda\sigma}.
\label{eq:fl}
\end{equation}
Where,
\begin{eqnarray}
 < \frac{\partial^2 {\cal L}}{\partial {\cal F}_{\mu\nu} 
\partial {\cal F}^{\lambda\sigma}}
  > &=& - \frac{(2 \pi \alpha')^2 T_3}{g}\left[- \frac{1}{4}\,
 (2\pi \alpha')^2 < {\cal F} \tilde{\cal F}> \, 
\epsilon^{\mu\nu}\,_{\lambda\sigma}
+ \left(1 - \frac{1}{4} \,(2 \pi \alpha')^2  
< {\cal F}^2>\right) \,\delta^{\mu\nu}_{\lambda\sigma}\right]
\nonumber\\
&\,&\quad + T_3 (2\pi \alpha')^2 \,\epsilon^{\mu\nu}\,_{\lambda\sigma},
\nonumber\\
&=& n \left[- \frac{1}{2 \pi g} \,\left(1 - \frac{q^2}{2 n^2} 
+ \frac{g^2 (p - q\chi_0)^2}{2 n^2}\right)\, 
\delta^{\mu\nu}_{\lambda\sigma} \right.\nonumber\\
&\,&\qquad \left.+ \frac{1}{2\pi} \left(\chi_0 + 
\frac{q (p - q\chi_0)}{n^2}\right)
\,\epsilon^{\mu\nu}\,_{\lambda\sigma}\right]. 
\label{eq:coef}
\end{eqnarray}
Note that since the background fields in [3] were obtained 
 upto the quadratic order in gauge fields, we have evaluated
$< \frac{\partial^2 {\cal L}}{\partial {\cal F}_{\mu\nu} \partial {\cal
F}^{\lambda\sigma}} >$ in the first expression of Eq.(4.16) upto that order. 
Also note that we have used $<{\cal F}^2> + <{\tilde {\cal F}}^2> = 0$ and
\begin{equation}
{\rm det} \left(\eta_{\mu\nu} + (2 \pi \alpha') {\cal F}_{\mu\nu}\right) 
= - \left[ 1 + \frac{1}{2} (2 \pi \alpha')^2 
{\cal F}^2 - \frac{1}{16} (2 \pi \alpha')^4 ({\cal F} 
\tilde{\cal F})^2 \right].
\end{equation}
It should be pointed out that in obtaining the second expression of Eq.(4.16), 
we have used Eq.\ (\ref{eq:ansatz}) and 
$T_3 = n T_0^3 = n /((2 \pi)^3 \alpha'^2)$.
 We therefore find the 
effective Lagrangian to this order to be of the form,
 \begin{equation}
 {\cal L}_f /n = - \frac{1} { 8 \pi g} \left(1  - \frac{q^2}{2 n^2} 
+ \frac{g^2 (p - q \chi_0)^2}{2
 n^2}\right) f_{\mu\nu} f^{\mu\nu} + \frac{1}{16 \pi} 
\left(\chi_0 + \frac{q (p - q\chi_0)}{n^2}\right) 
\epsilon^{\mu\nu \lambda\sigma} f_{\mu\nu} f_{\lambda\sigma},
\end{equation}
where the overall integer $n$ indicates that we have $n$ copies of the 
$U(1)$ Lagrangian. 
We can read, from the above action, the gauge coupling and the theta angle, 
 as
\begin{eqnarray}
\frac{1} {g^2_{\rm YM}} &= &\frac{1} {2 \pi g} \left(1 - \frac{q^2}{2 n^2} 
+ \frac{g^2 (p - q \chi_0)^2}{2
 n^2}\right), \nonumber\\
 \theta &=& 4 \pi \left(\chi_0 + \frac{q (p - q\chi_0)}{n^2}\right),
 \label{eq:cpt}
 \end{eqnarray}
 which, in the linear approximation, agree completely with the predictions of
 the $AdS_5 \times S^5$ correspondence, i.e., $g^2_{\rm YM} = 2 \pi g_{\rm eff}$
 and $\theta = 4\pi \chi_{\rm eff}$, respectively. We can examine this simply by
 expanding  $g_{\rm eff}$ given by 
Eq.\ (\ref{eq:esc}) and  $\chi_{\rm eff}$ given by Eq.\ (\ref{eq:chi}) to
quadratic order in charge $p$ and $q$ as:  
\begin{equation}
\frac{1}{g_{\rm eff}} = \left[ \frac{1}{g} - \frac{1}{2 g} 
\left(\frac{q}{n}\right)^2 + \frac{1}{2 g} 
\left( \frac{g (p - \chi_0 q)}{n}\right)^2\right],
\label{eq:gcoupling}
\end{equation}
and 
\begin{equation}
\chi_{\rm eff} = \chi_0 + \frac{q (p - q\chi_0)}{n^2}.
\end{equation}
So we confirm the predictions of the $AdS_5 \times S^5$ 
correspondence between the SYM coupling
and the effective string coupling and between the theta angle 
and the value of the RR
scalar evaluated at the horizon.
In analogy with the bulk picture of the F-strings and D-strings in the 
decoupling limit, 
the effects of the electric and magnetic
background fields are absorbed into the gauge coupling constant
$g_{\rm YM}^2 = 2 \pi g_{\rm eff}$ and the theta angle 
$\theta = 4 \pi \chi_{\rm eff}$
 and they do not appear in the resulting SYM theory.  
Therefore, we expect that the superconformal symmetry
is then restored. Thus we have a consistent picture both in the bulk 
and on the brane.

  Given the $AdS_5/CFT_4$ correspondence with respect to the equivalent
simple D3 brane,  we should have an $AdS_5/CFT_4$ correspondence with 
respect 
to the ((F, D1), D3) configuration with
the string coupling $g_{\rm eff}$ and the gauge coupling 
$g_{\rm YM}^2 = 2\pi g_{\rm eff}$. This in turn
indicates that Maldacena's $AdS_5/CFT_4$ correspondence holds true even 
for 
this non-trivial D3 brane configuration but now with the string coupling 
$g_{\rm eff}$ rather than the usual
string coupling $g$.

  As in the case of (F, D3) bound state, the effective string coupling
in Eq.(3.6) is again quantized but now in terms
of integers $p, q, n$. From Eq.\ (\ref{eq:esc}) we find that the F-strings 
in ((F, D1), D3) tend to decrease
the coupling while the D-strings tend to increase it. Therefore, 
we may control the string coupling in the
near-horizon region by adding or subtracting F-strings (or D-strings).

\section{T-dual Daughters of ((F, D1), D3)\protect\\}
  \label{sec:dnbs}

Once we have the space-time configuration for the non-threshold 
((F, D1), D3) bound state as given
in section 2, its descendants  can be 
obtained from this
by T-dualities, as described in detail in \cite{lurtwo}. T-dualities can
be applied either along the transverse directions of the D3 brane 
in ((F, D1), D3) or along the longitudinal directions
of the D3 brane which includes the direction of the (F, D1)-strings. 
Even though these
configurations can, in principle, be obtained directly from the 
equations of motion of type IIA or 
type IIB supergravity theory, it would be very difficult to obtain them 
in practice if we do not have these
symmetries in hand. There is no doubt that these configurations will play 
an important role in the
formulation of the unique M- or U-theory.

In this section, we  will first discuss all possible non-threshold 
bound states which can be obtained from
((F, D1), D3) by T-dualities\footnote{We do not consider any Dp brane 
resulting from D9 brane.}, 
following the table
given below:

\begin{center}
\begin{tabular}{|c|c|c|}  \hline
  & Parallel
 & Transverse\\
 \hline
Dp& D(p $-$ 1) & D(p + 1)\\
F& W  & F \\
W& F  & W \\
NS5& NS5 &KK \\
KK & KK & NS5 \\
\hline\end{tabular}
\end{center}

In this table W, F, NS5 and KK  denote waves, fundamental strings,  
NS fivebranes, and KK
monopoles, respectively, and they  are associated with NSNS fields.  
Dp ($ - 1\le p \le 8$)
are the so-called D-branes and are associated with  the RR fields.

Then we will present explicit Einstein-frame metric and dilaton for 
each of the non-threshold 
((F, Dp), D(p + 2)) bound states for $ 2 \le p \le 5$ where the 
well-known (Dp, D(p + 2)) bound states
are just special cases.

Let us begin with the T-dualities along the transverse directions 
of the D3 brane. We denote the coordinates
along the D3 brane as $x^0, x^1, x^2, x^3$, and $4, 5, 6, 7, 8, 9$ 
as its transverse directions. We also
denote $({\rm T}_i: \mapsto)$ as the T-duality along the $i$-th direction. 
According to the above table, we have
\begin{eqnarray}
&&(({\rm F}, {\rm D1}), {\rm D3}) \, ({\rm T}_4: \mapsto)\, (({\rm F}, 
{\rm D2}), {\rm D4})\, ({\rm T}_5: \mapsto)\, (({\rm F}, {\rm D3}), 
{\rm D5})\, ({\rm T}_6: \mapsto)\, 
(({\rm F}, {\rm D4}), {\rm D6})\nonumber\\
&&\, ({\rm T}_7: \mapsto)\, (({\rm F}, {\rm D5}), {\rm D7})\, 
({\rm T}_8: \mapsto)\, (({\rm F}, {\rm D6}), {\rm D8}).
\label{eq:tt}
\end{eqnarray}

We can T-dualize each of the above ((F, Dp), D(p + 2)) for $1 \le p \le 6$ 
along the longitudinal
directions of the original D3 brane, i.e., 1, 2 , 3 directions. We end 
up with different bound
states depending on whether we T-dualize along the `1' direction 
first or not. 
We demonstrate this on ((F, D1), D3) as examples. The other cases 
just follow. If we
T-dualize ((F, D1), D3) along the `1', i.e., the direction of 
(F, D1)-strings, we end
up with ((W, D0), D2). If we T-dualize along the `3', we have 
((F, D2), D2). We can now T-dualize on
((F, D2), D2) along either `2'
or `1'. In the former case, we have ((F, D3), D1) while for the 
latter we have ((W, D1), D1). We can 
further T-dualize ((F, D3), D1) along the `1', we end up with ((W, D2), D0).

In general, we have ((W, D(p $-$ 1)), D(p + 1)) if we T-dualize along 
`1' direction first on
the ((F, Dp), D(p + 2)) for $ 1 \le p \le 6$. If we T-dualize first 
along `3' instead, 
we have ((F, D(p + 1)), D(p + 1)). If we T-dualize on 
((F, D(p + 1)), D(p + 1)) along either `1' or
`2', we end up with ((W, Dp), Dp) or ((F, D(p + 2)), Dp), respectively. 
Further we can T-dualize on
((F, D(p + 2)), Dp) along `1', we end up with ((W, D(p + 1)), D(p $-$ 1)). 
We summarize all the bound states obtained this way in the
following table where the second column indicates the common directions 
shared by the two D branes in the
corresponding bound state:

\begin{center}
\begin{tabular}{|c|c|}  \hline
  Bound States
 & No. Common Dir.\\
 \hline
((W, D(p $-$ 1)), D(p + 1))& $p - 1$\\
((W, D(p + 1)), D(p $-$ 1))& $p - 1$ \\
((W, Dp), Dp)& $p  - 1$\\
((F, D(p + 1)), D(p + 1))& $p$\\
((F, D(p + 2)), Dp)& $p$\\
((F, Dp), D(p + 2))& $p$\\
\hline\end{tabular}
\end{center}

These are all possible non-threshold bound states which can be obtained 
from ((F, D1), D3) by T-dualities. 
Among these bound states, the well-known (Dp, D(p +2)) bound states  
for $ 0 \le p \le 6$ 
\cite{pol} appear as special cases when 
the charges associated with either F-strings or W-waves are set to zero. 
If we set the
charge associated with the W-waves in ((W, Dp), Dp) or the F-strings 
in ((F, D(p + 1)), D(p + 1)) for 
$ 1 \le p \le 6$ to zero, we end up with non-threshold 
(Dp, Dp) bound states $1 \le p \le 7$ which appear to be new. 
As mentioned above, the two Dp's share only 
$(p -1)$ common directions.
For those
in the type IIB theory, we can apply SL(2,Z) to obtain new bound states. 
For examples,
from ((F, D3), D3), we can have (((F, D1), D3), D3). One can also 
perform the consistency checks of the
$AdS_5/CFT_4$ correspondence for these two bound states. 

If we T-dualize 
(((F, D1), D3), D3) along the 
common direction of the two D3's only, we have (((F, D2), D2), D2). 
If we now set the charge associated with
the F-strings to zero, we end up with a non-threshold bound state 
(D2, D2, D2). The three D2's share only one common
direction. We can obtain (D1, D1, D1) by T-dualizing along 
the common direction
 of the three D2's or we can obtain(Dp, Dp, Dp) for 
$ 3 \le p \le 7$ by T-dualizing
  along directions transverse to (D2, D2, D2). 
The three Dp's in (Dp, Dp, Dp) for 
  $1 \le p \le 7$ share $p -1$ common directions. 
We can also obtain (((F, D3), D3), D3) by
  T-dualizing along one of directions transverse
to (((F,D2), D2), D2). Then we apply SL(2,Z) to 
(((F, D3), D3), D3), we can obtain
((((F, D1), D3), D3), D3). If we T-dualize this state along 
the common direction of the three
D3's only, we obtain ((((F, D2), D2),D2),D2). By setting the 
charge associated the F-strings
to zero, we  end up with the non-threshold bound state 
(D2, D2, D2, D2). We  repeat what we
have done for (D2, D2, D2). We can have non-threshold bound states 
(Dp, Dp, Dp, Dp) for 
$1 \le p \le 6$ where the four Dp's share $p - 1$ common directions. 
We can continue this process
to have bound states (Dp, Dp, Dp, Dp, Dp) for $1 \le p \le 5$ and so on. 
Finally, in this way we can have a seemingly unique bound state 
(D1, D1, D1, D1, D1, D1, D1, D1, D1) where the nine D1's share no 
common direction.

Also, from ((F, D5), D5), we can have 
(((F, D1), (NS5, D5)), (NS5, D5)) by S-duality of type IIB theory.  
As we will demonstrate in a bit 
simpler case of this, i.e.,
((F, D1), (NS5, D5)), in \cite{lurfour} that the charges for the 
D5- and NS5-branes are  related to 
the charges for the F-strings and D-strings. This may suggest 
that the formation 
of the SL(2,Z) fivebrane bound states
\cite{schone,wittwo,lurpone} be closely related to that of (F, D1) bound states or 
$(m,n)$-strings. 
{}From ((F, D3), D5),
we can have 
(((F, D1), D3), (NS5, D5)) and so on. 

Once we have these new bound states, 
we can again apply T-dualities
to obtain other new bound states, then S-dualities on those in type IIB. 
We can continue this process to obtain
all possible non-threshold bound states by S- and T-dualities simply from 
the original (F, D1)-strings. All these non-threshold bound states preserve
one half of the spacetime supersymmetries.
Even though we are, at this point, unable to count how many of these bound 
states are there, we believe that the total
number of these bound states is finite and may be related to the 
number of generators of the largest 
finite U-duality group \cite{HT} in type II theory, i.e., $E_{8(+8)}$. After all, 
these bound states are obtained by 
repeated
applications of the S- and T-dualities. We speculate here that these 
bound states will form the multiplets of the
$E_{8(+8)}$ U-duality  symmetry in the yet unknown M- or U-theory. 
We wish to come back to
provide more evidence for this elsewhere.

 	In closing this section we provide, as examples, the explicit 
Einstein-frame metric and the dilaton for 
each of  the ((F, D(p $-$ 2)), Dp) bound states for $ 3 \le p \le 7$. 
The metric is
\begin{eqnarray}
d s^2 =&& e^{- \phi_0/2}  (H H')^{1/4} {H''}^{(p - 3)/8} 
\left[ H^{- 1} \left( - (d x^0)^2 + (d x^1)^2\right) +
H'^{-1} \left( (d x^2)^2 + (d x^3)^2\right)\right. \nonumber\\
&& \left. + {H''}^{-1} \left( (d x^4)^2 + \cdots + 
(d x^p)^2 \right) + dy^i dy^i \right],
\label{eq:nbsm}
\end{eqnarray}
with $i = 1, \cdots, 9 - p$; and the dilaton is
\begin{equation}
e^\phi = e^{\phi_0} (H' H)^{-1/2} {H''}^{(7 - p)/4}.
\label{eq:nbsd}
\end{equation}
In the above, we have chosen the string-frame metric to be 
asymptotically Minkowski. The  Harmonic function
$H$ is
\begin{equation}
H= \left\{\begin{array}{l}
           1 + \frac{Q_p}{r^{7 - p}}, \qquad  3 \le p \le 6,\\
	   1 - Q_7\, {\rm ln}\,r, \quad p = 7, \end{array}\right. 
\label{eq:hfh}
\end{equation}
with $Q_p = \Delta^{1/2}_{(p, q, n)} e^{3 \phi_0/2} 
\sqrt{2} \kappa_0 /[(7 - p) \Omega_{8 - p}]$ and the 
Harmonic functions $H'$ and $H''$ are
\begin{eqnarray}
H' &=& \left[(n^2 H + q^2)\, e^{-\phi_0}  + (p - \chi_0 q)^2\, 
e^{\phi_0}\right] \Delta^{-1}_{(p,q,
n)},\nonumber\\
H'' &=& \frac{q^2 \,H\, e^{- \phi_0} + (p - \chi_0 q)^2 \, 
H'\, e^{\phi_0}}{q^2\, e^{- \phi_0} + 
(p - \chi_0 q)^2\, e^{\phi_0}}.
\label{eq:nhfh'h''}
\end{eqnarray}
Other fields can be obtained from those given in section 3 for the 
((F, D1), D3) configuration, following the 
prescription given in \cite{lurtwo}. In the above, the 
$\Delta$-factor continues to be given by 
Eq.\ (\ref{eq:nndelta}) but now the  integer $p$ represents the 
quantized charge per $(2 \pi)^{p -1} \alpha'^{(p -1)/2}$
area over the $x^2 \cdots x^p$-plane carried by the F-strings, the integer 
$q$ represents the quantized charge per 
$(2 \pi)^2 \alpha'$ area over the $x^{p -1} x^p$-plane carried 
by the D(p $-$ 2) branes, and the integer $n$ is
the quantized (p + 2)-form charge carried by the Dp brane. 
The tension for each of these bound states can 
be calculated by generalizing the mass per unit $p$-brane volume given 
in \cite{lu} in a similar fashion
as we did in \cite{lurtwo}. The result is
\begin{equation}
T_p (p, q, n) = \frac{T_0^p}{g} \sqrt{n^2 + q^2 + g^2 (p - \chi_0 q)^2},
\label{eq:pt}
\end{equation}
where the subscript `$p$' in $T_p$ and the superscript `$p$' in $T^p_0$ 
should not be confused with the
integral charge $p$ for the F-strings.

\section{Conclusion\protect\\}
\label{sec:c} 

To conclude, we have constructed in this paper the explicit space-time 
configuration for the non-threshold ((F, D1), D3) bound state using
the type IIB SL(2,Z) symmetry and the known (F, D3) configuration. 
We have shown that the 
((F, D1), D3) bound state is identical to the bound state consisting 
of D3 branes carrying
quantized constant electric and magnetic fields. The $(p,q)$-strings 
are identical to the
corresponding electric and magnetic, or dyonic, flux lines. This also connects 
the type IIB SL(2,Z)
symmetry in the bulk with the SL(2,Z) symmetry in the SYM theory on the brane.
We have  studied the decoupling limit of this
bound state. We found that given the $AdS_5/CFT_4$ correspondence of
Maldacena with respect to a simple
D3 brane and the usual string coupling, we should have a similar 
correspondence with respect to 
this bound state but now with an effective string coupling.
This, in turn, indicates that Maldacena's $AdS_5/CFT_4$ correspondence 
holds true even for
a non-trivial D3 brane configuration with the corresponding 
effective string coupling rather 
than the usual string coupling. We also found that D-strings 
in ((F, D1), D3) bound state tend to
increase the effective string coupling while the F-strings tend to 
decrease it. This property
may be used to control the string coupling in the bulk and the 
SYM coupling on the brane in the near-horizon
region.

We  have given the list of all possible descendants  of this ((F, D1), D3) 
bound states by
T-dualities. Unless there is an obstacle preventing us to perform 
T-dualities and the type IIB SL(2,Z) on these bound states, we  have predicted
the existence of new non-threshold bound states (Dp, Dp) for $1 \le p \le 8$ 
where
the two Dp's share only $(p - 1)$ common dimensions, (Dp, Dp, Dp) for
$1 \le p \le 7$ with the three Dp's sharing $p - 1$ common directions,
(Dp, Dp. Dp, Dp) for $1 \le p \le 6$ with the four Dp's sharing $p - 1$
common directions and so on.  This process stops at the seemingly unique
bound state (D1, D1, D1, D1, D1, D1, D1, D1, D1) with the nine D1's sharing
no common direction.
Moreover, we  have presented the explicit configurations for the 
((F, Dp), D(p + 2)) 
bound states for $2 \le p \le 5$. Using the type IIB SL(2,Z) symmetries 
on those newly
obtained bound states, we can construct further complicated bound states. 
Then by T-dualities, we can
have even more bound states. We can probably exhaust all possible  
bound states by continuing this process.
All these bound states are BPS ones, preserving one half of the 
spacetime supersymmetries. 
These bound states can, in principle, be obtained from the equations of 
motion without the need to
use the S- and T-dualities. They are expected to appear in the 
uncompactified M- or U-theory.
We have conjectured that the total number of these bound states are 
finite and  these bound states will
form multiplets of the U-duality group $E_{8(+8)}$ in the yet unknown
M- or U-theory.

\acknowledgments
JXL acknowledges the support of NSF Grant PHY-9722090.	We thank 
Juan Maldacena for an 
e-mail correspondence.

\end{document}